\documentclass[aps]{revtex4}
\begin{document}
\renewcommand{\thefootnote}{\arabic{footnote}}
\setcounter{footnote}{0}
\vskip 0.4cm
\def\e{{\rm e}}
\def\ln{{\rm ln}} 
\title{\bf 
The Information Geometry\\ 
of the Ising Model\\
on Planar Random Graphs
}
\author{ 
W. Janke$^{a}$,
D.A. Johnston$^{b}$ 
and
Ranasinghe P.K.C. Malmini$^{c}$
}
\affiliation{
$^{a}$ {\it Institut f\"ur Theoretische Physik,
Universit\"at Leipzig,
Augustusplatz 10/11, 
D-04109 Leipzig, Germany}
\\
$^b$ {\it Department of Mathematics,
Heriot-Watt University,
Riccarton,
Edinburgh, EH14 4AS, Scotland}
\\
$^{c}$ {\it
Department of Mathematics,
University of Sri Jayewardenepura,
Gangodawila, Sri Lanka
}
}
\date{\today}
%\maketitle
%-----------------------------------------------------------------------
                      \begin{abstract}
%-----------------------------------------------------------------------
%
It has been suggested that an information geometric view of statistical
mechanics in which a metric is introduced onto the space of parameters
provides an interesting alternative characterisation
of the phase structure, particularly in the case where there are
two such parameters --
such as the 
Ising model with inverse temperature $\beta$ and external field $h$.

In various two parameter
calculable models the scalar curvature ${\cal R}$ of the 
information
metric has been found to diverge at the phase
transition point $\beta_c$ and a plausible
scaling relation postulated: ${\cal R} \sim |\beta- \beta_c|^{\alpha - 2}$.  For spin models the necessity of calculating
in non-zero field has limited analytic consideration to 1D,
mean-field and Bethe lattice Ising models.
In this paper we use the solution in field of the Ising model 
on an ensemble of planar random graphs (where $\alpha=-1, \; \beta=1/2,
\gamma=2$) to evaluate
the scaling behaviour of
the scalar curvature, and find ${\cal R} \sim | \beta- \beta_c |^{-2}$.  The apparent discrepancy
is traced back to the effect of a negative $\alpha$. 
 
%
%-----------------------------------------------------------------------
                        \end{abstract} 
%-----------------------------------------------------------------------
%
\maketitle
%
%***********************************************************************
%
%
%-----------------------------------------------------------------------
                  \pagenumbering{arabic}
%-----------------------------------------------------------------------

\section{Generalities: The Information Geometry}

Various authors, motivated by ideas
in parametric statistics \cite{Fish},
have discussed
the advantages of taking a geometrical perspective on statistical mechanics
\cite{Rupe,Jany,jany,Brody,Brian,BrianA,Brianetc}. The ``distance'' between two probability
distributions in parametric statistics can be measured using a geodesic distance which is calculated
from the Fisher information matrix for the system. 
In a statistical mechanical context the
probability distributions of interest are
Gibbs measures
\begin{equation} 
p(x|\theta)\ =\ \exp\left( -\sum_{i=1}^{r}\theta^{i}H_{i}(x) 
- \ln Z(\theta)\right)\ , 
\label{eq:gibbs} 
\end{equation} 
where the $x$ characterise the state of the system (e.g. spins),
the $H_i(x)$ are the various terms in the Hamiltonian,
the $Z(\theta)$ is the normalising partition function and
the $\theta^i$ are the various parameters such as the
inverse temperature $\beta$, the external field $h$, etc. 

The manifold ${\cal M}$ of parameters is endowed
with
a natural
Riemannian metric, the Fisher-Rao metric \cite{Fish},
which measures the distance between different configurations.
For a spin model in field ${\cal M}$ 
is a two-dimensional manifold parameterised by
$(\theta^{1},\theta^{2})=(\beta,h)$. The components of 
the  Fisher-Rao metric take the simple form $G_{ij} =  \partial_{i}\partial_{j} f $ in this case, where $f$ is the reduced free energy per site and
$\partial_{i} = \partial/\partial\theta^{i}$. A natural object to consider
in any geometrical approach is the scalar or Gaussian curvature
which may be calculated
as
\begin{equation} 
{\cal R}\ =\ - \frac{1}{2 G^{2}} 
\left| \begin{array}{lll} 
\partial^{2}_{\beta}     f & \partial_{\beta}\partial_{h}    f & 
\partial_{h}^{2}    f \\ 
\partial^{3}_{\beta}    f & \partial_{\beta}^{2}\partial_{h}    f & 
\partial_{\beta}\partial_{h}^{2}    f \\ 
\partial^{2}_{\beta}\partial_{h}    f & 
\partial_{\beta}\partial_{h}^{2}    f & \partial_{h}^{3}    f  
\end{array} \right| \ , 
\label{equcurv} 
\end{equation}
where $G={\rm det}(G_{ij})$. 

It is worth remarking that, unlike most
standard statistical mechanical observables, the 
curvature ${\cal R}$ depends
on third order derivatives. Nonetheless, a plausible
scaling relation has been advanced for ${\cal R}$ in the 
critical region. The hypothesis, on dimensional grounds,
is that the curvature
depends on the correlation volume 
for a second-order transition
${\cal R} \sim \xi^d$, where $\xi$ is the correlation length
and $d$ is the dimension of the system. 
This is reasonable since $\xi$ is the only physical scale in the
system near criticality.

Combined with
hyperscaling, $\nu d = 2 - \alpha$, and standard
scaling assumptions this leads to
\begin{equation}
{\cal R} \sim |\beta - \beta_c |^{  \alpha - 2}\ .
\label{equscal}
\end{equation}
In the above $\alpha$ is the standard 
exponent characterising the scaling of the specific heat,
so consideration of ${\cal R}$ clearly offers a way
of determining critical exponents in  a non-standard manner.

Analytic determination of ${\cal R}$ in spin models has been limited
by the necessity of carrying out calculations in field. One case
where this is possible is the 1D Ising model \cite{Jany}, where
the curvature was calculated to be
\begin{equation}
{\cal R}=1 + \eta^{-1}\cosh h
\end{equation}
with
$\eta=\sqrt{\sinh^{2}h+e^{-4\beta}}$. The 1D Ising model
can be thought of as having a zero-temperature transition, so
looking at $h=0$, $\beta \rightarrow \infty$ we see that
${\cal R} \sim e^{2\beta}$, corresponding
to the expected $\alpha=1$. Similarly, it is possible
to obtain an expression for the scalar curvature
for the Ising model on a Bethe lattice \cite{Brian},
where the scaling behaviour is also verified. Both of these
examples have unsatisfactory aspects -- the 1D Ising model
has no real transition and the Bethe lattice Ising model
is mean field in nature.

Given the relative paucity of models which
are soluble in field
any further explicit calculations
would be welcome, particularly in a non-mean-field
model with a genuine finite-temperature phase transition.
In the sequel we discuss one such case, the Ising model
on dynamical planar random graphs.

\section{Particulars: The Ising Model on Planar Graphs}

The solution of the Ising model on 
an ensemble of $\Phi^4$ ($4$-regular) or $\Phi^3$
($3$-regular)
planar random graphs was first presented  
by Boulatov and Kazakov \cite{kaz,BK},
who were motivated by string-theoretic
considerations, since the 
continuum limit of the theory represents
matter coupled to 2D quantum gravity. They considered the partition function
for the Ising model on a single $n$ vertex planar graph 
with connectivity matrix $ G_{ij}^n $ 
\begin{equation}
Z_{{\rm single}}(G^n,\beta,h) =
\sum_{\{\sigma\}} \exp \left({\beta}\sum_{\langle i,j \rangle} G^n_{ij}\sigma_i
\sigma_j + h \sum_i \sigma_i\right)\, ,
\end{equation} 
then summed it over all $n$ vertex graphs $\{G^n\}$ 
resulting in
\begin{equation}
Z_n = \sum_{\{G^n\}} Z_{{\rm single}}(G^n,\beta,h)\ ,
\end{equation} 
before finally forming the grand-canonical sum over 
graphs with different numbers $n$ of
vertices 
\begin{equation}
{\cal Z} = \sum_{n=1}^{\infty} \left( - 4 g c \over 
( 1 - c^2 )^2 \right)^n Z_n\ ,
\label{grand}
\end{equation}
where $c = \exp ( - 2 \beta)$.   
This last expression
could be calculated exactly as matrix integral
over $N \times N$ Hermitian matrices,
\begin{equation}
{\cal Z} = - \log \int {\cal D}\phi_1~{\cal D}\phi_2~ 
\exp \left( -{\rm Tr}\left[{1\over 2}(\phi_1^2+\phi_2^2)- c \phi_1\phi_2  - 
\frac{g}{4}( \e^h \phi_1^4 + \e^{-h} \phi_2^4)\right]  \right),
\label{matint}
\end{equation}
where the $N \to \infty$ limit is to be taken to pick out the planar diagrams
and the potential appropriate for $\Phi^4$ (4-regular) random 
graphs has been shown.

When the matrix integral is carried out the solution is given 
in parametric form by
\begin{equation}
{\cal Z} = {1\over 2}\log {z \over g}-{1\over g}\int_0^z~{dt\over t}g(t)
+{1\over 2g^2}\int_0^z{dt\over t}g(t)^2,
\label{fullpart}
\end{equation}
where the function $g(z)$ is 
\begin{equation}
g(z)=\frac{1}{9} c^2 z^3 +  \frac{z}{3}
\left[ \frac{1}{(1- z)^2} - c^2 +\frac{z B}{(1- z^2)^2}
\right]
\label{geq}
\end{equation}
and $B= 2 [ \cosh ( h ) - 1]$.

In the thermodynamic limit the free energy per site is given by 
\begin{equation}
f  
=  \log \left( {-4 c g (z_0) \over ( 1 - c^2)^2} \right)\ ,
\label{equfL}
\end{equation}
where $z_0 = z_0 ( \beta, h)$ is the appropriate low- or high-temperature 
solution of
$ g' ( z ) =0$. When $h=0$ this may be solved in closed form, and although the
solution is not available explicitly
for non-zero $h$ it can still be developed
as a power series in $h$ around the zero-field solutions in order to 
obtain expressions for quantities such as the energy, specific \ heat,
magnetization and
susceptibility. It was found that
the critical exponents were given by $\alpha=-1$, $\beta=1/2$, $\gamma=2$,
so the transition was {\it third} order with, intriguingly, the same
exponents as the 3D spherical model on a regular lattice \cite{Stau}.

If we carry out a perturbative expansion
for the high-temperature solution,
which is symmetric in $h$ and hence a series in even powers, we
find
\begin{eqnarray}
z_{0} &=&
1-\frac{1}{u}-\frac{(u-1)(2 u^2-2 u+1)}{(2 u-1)^4} h^2 \nonumber \\
&+& \frac{(u-1) (2 u^2-2 u+1) ( 4 u^5-10 u^4+10 u^3-5 u^2+5 u+1)}{24 (2 u-1)^9} h^4 + \dots\ ,
\label{equz0}
\end{eqnarray}
where the coefficients
in the series are
 most naturally expressed in terms of $u = \exp ( - \beta)
= \sqrt{c}$, as above. 

\section{Generalities: Scaling of The Scalar Curvature}

The expected scaling form of the various components
of ${\cal R}$ for a 
generic spin model in field is discussed at some length in \cite{jany},
and we now recapitulate these results briefly for comparison with
the specific results for the Ising model on planar random graphs
in the next section. The starting point is the scaling form
of the free energy per spin near the critical point,
\begin{equation}
f(\epsilon, h) = \lambda^{-1} f ( \epsilon \lambda^{a_{\epsilon}} , 
h \lambda^{a_h} )\ ,
\end{equation}
where $\epsilon \equiv \beta_c - \beta$ and $a_{\epsilon}, a_h$
are the scaling dimensions for the energy and spin operators. For
$\epsilon>0$, i.e., in the unbroken high-temperature phase, we can use 
standard scaling assumptions to write this as
\begin{equation}
f(\epsilon, h) =  \epsilon^{1 / a_{\epsilon}} \psi_{+} ( h \epsilon^{- a_{h} / a_{\epsilon}} ),
\end{equation}
where $\psi_{+}$ is a scaling function
and we also define $A = 1 / a_{\epsilon}$ and $C = - a_{h} / a_{\epsilon}$
for later convenience.  In terms of the standard critical exponents
$A = 2 - \alpha$ and $A + C = \beta$.

This generic scaling form can now be substituted into
equ.~(\ref{equcurv}) to find the scaling behaviour 
of the various components and the 
scalar curvature (\ref{equcurv}) itself near criticality (i.e. $h=0$, 
$\epsilon \rightarrow 0$),
\begin{equation}
{\cal R} = -\frac{1}{2 G^{2}}
\left| \begin{array}{ccc}
A (A - 1) \epsilon^{A-2} \psi_{+} ( 0 ) & 0  &
\epsilon^{A + 2 C} \psi_{+}^{''} ( 0 ) \\
\!\!-A (A - 1) (A -2 ) \epsilon^{A-3} \psi_{+} ( 0 ) & 0  &
 \!\!\!\!-(A + 2 C ) \epsilon^{A + 2 C -1 } \psi_{+}^{''} ( 0 )\! \\
0 &
\!\!\!\!-(A + 2 C ) \epsilon^{A + 2 C -1 } \psi_{+}^{''} ( 0 ) & \epsilon^{A + 3 C} \psi_{+}^{'''} ( 0 )
\end{array} \right|\ , 
\label{equcurv2}
\end{equation}
where the scaling of the metric determinant is
\begin{equation}
G = A (A - 1) \epsilon^{ 2 A + 2C -2 } \psi_{+} ( 0 ) \psi_{+}^{''} ( 0 )\ .
\end{equation}
Expanding the determinant one finds two terms of similar order
contributing to give
\begin{equation} 
{\cal R } =  { ( A + 2 C ) [ (A + 2 C ) - ( A - 2 )] \over 2 A ( A -1 ) \psi_{+} ( 0 ) }  \epsilon^{-A}
\end{equation}
or, translating back to the standard critical exponents,
\begin{equation}
{\cal R } = { \gamma ( \gamma - \alpha ) \over 2 ( 2 -\alpha ) ( 1 - \alpha ) \psi_{+} ( 0 )}
\epsilon^{\alpha -2}\ .
\end{equation}
The discussion in \cite{jany} was intended to be as general as possible,
one should note that for Ising-like models with a $\pm h$ symmetry
all odd derivatives of the scaling function w.r.t.\ $h$ will vanish
so $ \partial^3_h f =0$ rather than $\epsilon^{A + 3 C} \psi_{+}^{'''} ( 0 )$.
This does not affect the stated scaling relations. 

However, one feature
of these scaling relations does have an impact on the observed
scaling for the Ising model. Generically one expects that
$\partial_{\beta}^2 f = A (A - 1) \epsilon^{A-2} \psi_{+} (0)$,
which contributes to both the metric and the determinant
involved in calculating ${\cal R}$. If $A>2$, i.e. $\alpha<0$, this
naively suggests a vanishing $\partial_{\beta}^2 f$ at
criticality, which will in general {\it not} be the case. 
There would instead be a  contribution from a regular term, which would 
give a constant at the critical point. Having such a 
constant term modifies the scaling form of ${\cal R}$ in the case $\alpha < 0$,
$A > 2$ to
\begin{equation}
{\cal R} = -\frac{1}{2 G^{2}}
\left| \begin{array}{ccc}
A ( A - 1) \phi ( 0) & 0  &
\epsilon^{A + 2 C} \psi_{+}^{''} ( 0 ) \\
\!\!-A (A - 1) (A -2 ) \epsilon^{A-3} \psi_{+} ( 0 ) & 0  &
\!\!\!\!-(A + 2 C ) \epsilon^{A + 2 C -1 } \psi_{+}^{''} ( 0 )\! \\
0 &
\!\!\!\!-(A + 2 C ) \epsilon^{A + 2 C -1 } \psi_{+}^{''} ( 0 ) & \epsilon^{A + 3 C} \psi_{+}^{'''} ( 0 )
\end{array} \right|\ ,
\label{equcurv3}
\end{equation}
where we have denoted the constant by $A ( A - 1) \phi ( 0)$.
The scaling for $G$ is also modified to
\begin{equation}
G = A (A - 1) \epsilon^{ A + 2C} \phi ( 0 ) \psi_{+}^{''} ( 0 )\ .
\label{equG3}
\end{equation}

When expanded the expression
for ${\cal R}$ contains two term which now have differing orders
in $\epsilon$. The leading term for $A>2$, the case which we are interested
in, is
\begin{equation}
{\cal R}\ =   { ( A + 2 C )^2 \over 2 A ( A -1 ) \phi ( 0 ) }  \epsilon^{-2}
\label{equR3}
\end{equation}
or
\begin{equation}
{\cal R } = { \gamma^2 \over 2 ( 2 -\alpha ) ( 1 - \alpha ) \phi ( 0 )}
\epsilon^{ -2}\ ,
\label{equR3a}
\end{equation}
so the critical exponent $\alpha$ no longer appears in the scaling
exponent.

By virtue of the 
Boulatov and Kazakov solution the Ising model on planar random graphs allows
us to explicitly confirm these observations, as we see in the next section.
Since $\alpha=-1, \beta = 1/2, \gamma = 2$, we have $A=3$, $C=-5/2$
and the modified discussion of scaling should apply.

\section{Particulars: The Scalar Curvature for Ising}

We can now take the series expansion for $z_0$ from  equ.~(\ref{equz0}),
insert this into $g(z)$ and 
use the resulting expression for $f$
in  equ.~(\ref{equfL}) 
to calculate the various terms
that appear in the scalar curvature ${\cal R}$ 
as power series
in $h^2$.
We find that the leading terms at $h=0$, with $\epsilon_u \equiv u - u_{cr} =
\epsilon/2 + \ldots$
and $u_{cr} = 1/2$, and using $\beta, h$ as co-ordinates are
\begin{equation}
{\cal R}\ =\ - \frac{1}{2 G^{2}}
\left| \begin{array}{ccc}
{352 \over 225} & 0 & {3 \over 20} \epsilon_u^{-2} \\[0.1cm]
-{1072 \over 675} & 0  & {3 \over 20} \epsilon_u^{-3} \\[0.1cm]
0 &  {3 \over 20} \epsilon_u^{-3} &  0 
\end{array} \right|\ .
\label{equcurv4}
\end{equation}
The determinant of the metric scales as
$G = {88 \over 375 } \epsilon_u^{-2} + \ldots$ so the final scaling expression for
the scalar curvature is
\begin{equation}
{\cal R} \sim \frac{225}{704 } \epsilon_u^{-2} + \ldots = \frac{225}{176}
\epsilon^{-2} + \ldots\ .
\label{finalscaling}
\end{equation}
% A glance back at equs.~(\ref{equcurv3},\ref{equG3},\ref{equR3},\ref{equR3a}) 
A glance back at equs.~(\ref{equcurv3})--(\ref{equR3a}) 
shows that the modified scaling
for $A>2$ that these incorporate is, indeed, followed for the individual components
in equ.~(\ref{equcurv3}), the metric in  equ.~(\ref{equG3}) and the scalar curvature
itself in equs.~(\ref{equR3}) and (\ref{equR3a}).

It is an easy matter to calculate ${\cal R}$ for
any $u$ when $h$ is small,
using the expansion for $z_0$ in equ.~(\ref{equz0}).
Writing
\begin{equation}
{\cal R} = {\cal R}_0 + {\cal R}_2 h^2 + \ldots
\end{equation}
the first two coefficients are given by
\begin{eqnarray}
{\cal R}_0 &=&
\frac{(6 u^8+43 u^7+357 u^6+1265 u^5+2123 u^4+1841 u^3+783 u^2+75 u+3) 
}{2 (6 u^5+27 u^4+56 u^3+54 u^2+18 u+3)^2 (2 u
-1)^2} \nonumber \\
&\times& 
(2 u^2+2 u+1) (u+1) (u^2+1)
\end{eqnarray}
and
\begin{eqnarray}
{\cal R}_2 &=& \frac{1}{2} (u + 1) (u^{2} + 1)  u  (u + 2)
\left[ 144   u^{18} + 1008  u^{17} - 3276  u^{16} - 31180  u^{15} - 79106  u ^{14} \right.\nonumber \\
&-& 129786  u^{13} - 135424  u^{12} - 92093  u^{11} - 78645   u^{10} - 37499  u^{9} + 54941  u^{8} \nonumber \\
&+& 245658  u^{7} + 410788  u^{6} + 328760  u^{5} + 139986 u^{4} + 33183  u^{3} + 5331  u^{2} + 765  u \nonumber \\
&+& \left. 45 \right]/\left[(6  u^{5} + 27  u^{4} + 56  u^{3} + 54  u^{2} + 18   u + 3)^{3}  (2\,u - 1)^{7} \right]\ , 
\end{eqnarray}
which give the scaling of equ.~(\ref{finalscaling}) when $u$ is set equal 
to $1/2 + \epsilon_u$.

In Fig.~1 we have plotted ${\cal R}$ close to $u_{cr} = 1/2$
using a series correct up to $O(h^6)$ terms. 
The scaling region in $h$ is very narrow, with the approximation to ${\cal R}$
rapidly giving large negative values outside this region
due to the increasingly strong divergences
in the series coefficients as $u \rightarrow u_{cr}$ for
increasing order in $h$. This turnover is just visible on the edges
of the plotted surface.
The sensitivity
to $h$ would have to be carefully handled in any numerical investigations
of ${\cal R}$. Within the domain of validity of the expansion in $h$ it appears
that ${\cal R}$ is positive. It has been remarked \cite{Jany,Brody}
that ${\cal R}$ is positive in the thermodynamic
limit for Ising models when the parameters take physical
values, with the only divergence being at the critical point, and the Ising
model here provides another example. This feature is apparently not
universal, calculations of ${\cal R}$ for the one-dimensional Potts
model \cite{us} and field theories \cite{Brianetc} do not give positive 
curvatures throughout the physical parameter space.

It has been observed that the line $h=0$
is a
geodesic of the metric for the one-dimensional
Ising and Potts models \cite{us}.
The geodesic equations using co-ordinates
$\beta, h$ are given in general by
\begin{eqnarray}
{d V^\beta\over ds} + \Gamma^\beta_{\beta\beta}V^\beta V^\beta
+ 2\Gamma^\beta_{\beta h}V^\beta V^h + \Gamma^\beta_{h h }V^h V^h
& = & \lambda(s) V^\beta\ , \label{geodesic1} \\
{d V^h\over ds} + \Gamma^h_{\beta\beta}V^\beta V^\beta
+ 2\Gamma^h_{\beta h}V^\beta V^h + \Gamma^h_{h h }V^h V^h 
& = & \lambda(s) V^h\ ,
\label{geodesic2}
\end{eqnarray}
where $s$ parameterises the flow
lines, $V^\beta = {d \beta / ds}$, $V^h = {d h / ds}$,
the $\Gamma$ are the Christoffel symbols
and $\lambda(s)$ allows for the possibility of a non-affine parameter
choice.

A vector field with a flow line along $h=0$
has $V^h=0$, so in this case equs. (\ref{geodesic1}) and (\ref{geodesic2})
reduce to 
\begin{eqnarray}
{d V^\beta\over ds} + 
\left(\Gamma^\beta_{\beta\beta}\right)\vrule height14pt depth6pt\raise -4pt \hbox{$_{h=0}$}V^\beta V^\beta
&=&\lambda(s) V^\beta\ , \\
\left(\Gamma^h_{\beta\beta}\right)\vrule height14pt depth6pt\raise -4pt \hbox{$_{h=0}$}V^\beta V^\beta
& = & 0\ .
\end{eqnarray}
The first of these equations for 
$\beta(s)$ always
has a solution and the second requires
$\Bigl(\Gamma^h_{\beta\beta}\Bigr)\vrule height12pt depth4pt\,\raise -2pt \hbox{$_{h=0}$}=0$.
This is satisfied for the Ising model on planar random graphs
because
the Christoffel symbol $\Gamma^h_{\beta\beta}$ vanishes at
$h=0$ for the same reason as in the one-dimensional models --
the off-diagonal components of the metric $\partial_{\beta} \partial_h f$
are $O(h)$ in all cases. We therefore
find that for the Ising model on planar
random graphs $h=0$ is also a geodesic line in the $\beta, h$ plane.

We close with a remark on the infinite temperature ($T\rightarrow \infty$, 
$\beta \rightarrow 0$, $u \rightarrow 1$) limit of
${\cal R}$ when $h=0$. In this limit ${\cal R}$ was found to
be 2 for the one-dimensional Ising model \cite{Jany} and
$z/2$ for the Ising model on a $z$ co-ordinated Bethe lattice
\cite{Brian}. Here we find that ${\cal R} ( T = \infty) = {4060 \over 1681} =
2.415....$, so if we accept the suggestion  in \cite{Jany}
that ${\cal R} - {\cal R} ( T = \infty)$ should be taken as a measure
of fluctuations caused by the spin interactions the correct measure
of the deviation from ideal paramagnetism 
for the Ising model on planar random graphs
is ${\cal R} - {4060 \over 1681}$.

\section{Conclusions}

We have calculated the scaling behaviour 
of the scalar curvature of the Fisher-Rao metric
for the Ising model on planar
random graphs using the exact solution of \cite{kaz,BK}
combined with a perturbative expansion in the external field $h$.
Although $\alpha=-1$ for this model, ${\cal R} \sim \epsilon^{-2}$
rather than the ${\cal R} \sim \epsilon^{-3}$ one might have
expected naively from general
scaling arguments. This discrepancy
was traced back to the effect that a negative value of $\alpha$
had on the scaling of the various components of the metric and the
terms which contributed to ${\cal R}$.

Various qualitative features of the calculated  ${\cal R}$ tally with
earlier observations of one-dimensional and mean-field Ising models.
It is positive (within the
domain of applicability of our semi-perturbative calculation)
and diverges only at the critical point. The zero-field line is seen to
be a geodesic, just as for the one-dimensional Ising and Potts models.

It would be an
interesting exercise to calculate
${\cal R}$ for other models where some form
of  solution in field was accessible.

\section{Acknowledgements}

W.J. and D.J. were partially supported by
EC IHP network
``Discrete Random Geometries: From Solid State Physics to Quantum Gravity''
{\it HPRN-CT-1999-000161}.

\bigskip
%

%------------------------------------

\clearpage \newpage
\begin{figure}[t]
\vskip 15.0truecm
\includegraphics{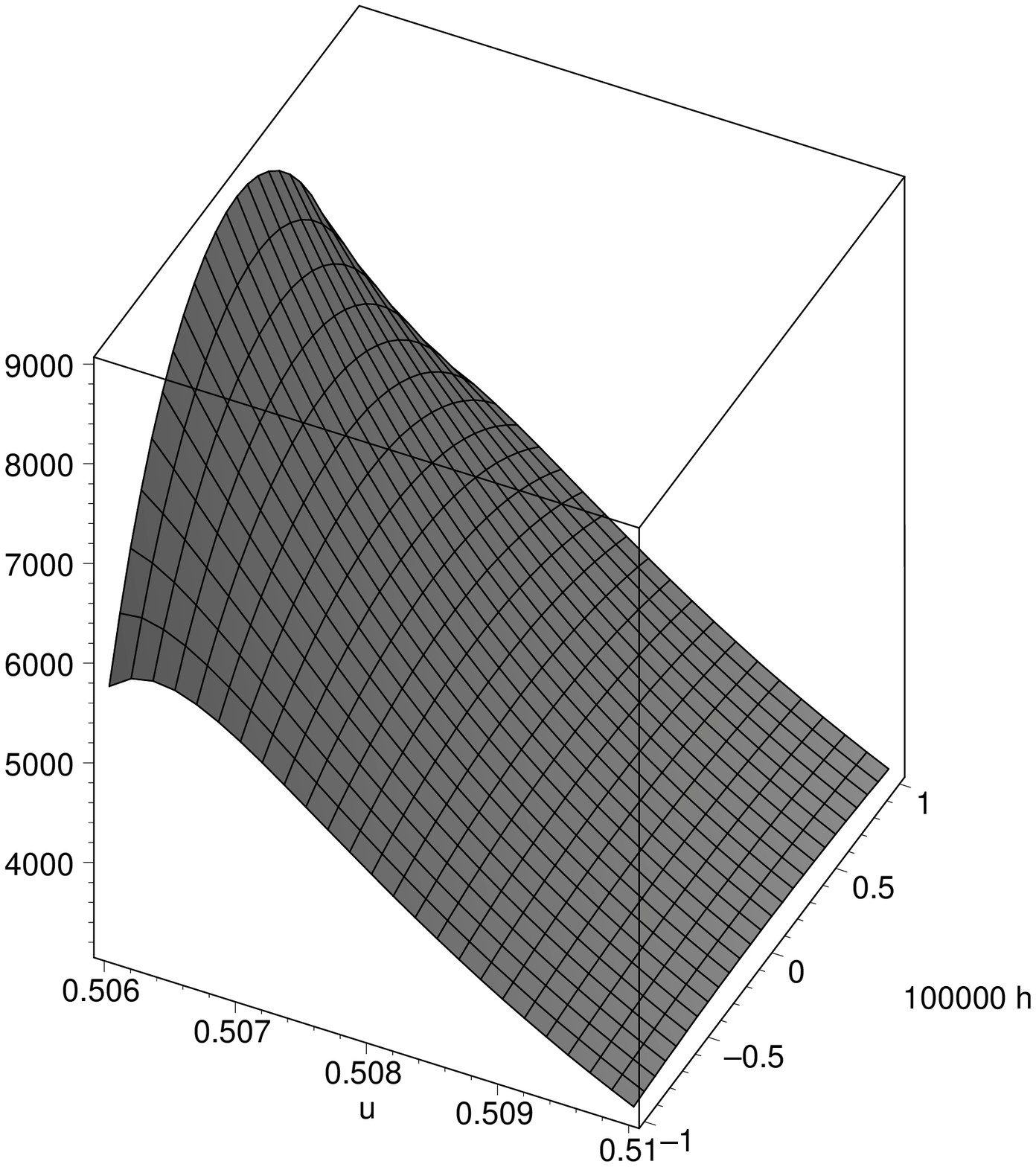}
\caption[]{\label{fig3} A plot of ${\cal R}$ close to $u_{cr}=1/2$.
Note that the external field $h$ is scaled by a factor of $10^5$ and so covers a very narrow range.
}
 
\end{figure}
 
%------------------------------------

\end{document}